# Detection and Prevention of Botnets and malware in an enterprise network

*Manoj Rameshchandra Thakur, Divye Raj Khilnani, Kushagra Gupta, Sandeep Jain and Vineet Agarwal

Computer Science Department, V.J.T.I., Mumbai, India.
E-mail: manoj.thakur66@gmail.com
E-mail: divyeraj88@gmail.com
E-mail: kush234@gmail.com
E-mail: sandeepjain88@gmail.com
E-mail: vineet140@gmail.com
*Corresponding Author

Suneeta Sane

Computer Science Department,
V.J.T.I, Mumbai, India.
E-mail: sssane@vjti.org.in

Sugata Sanyal

Tata Institute of Fundamental Research
Mumbai, India.
E-mail: sanyal@tifr.res.in

Prabhakar S. Dhekne

Bhabha Atomic Research Centre
Mumbai, India.
E-mail: dhekne@barc.gov.in

**Abstract**: One of the most significant threats faced by enterprise networks today is from Bots. A Bot is a program that operates as an agent for a user and runs automated tasks over the internet, at a much higher rate than would be possible for a human alone. A collection of Bots in a network, used for malicious purposes is referred to as a Botnet. Bot attacks can range from localized attacks like key-logging to network intensive attacks like Distributed Denial of Service (DDoS). In this paper, we suggest a novel approach that can detect and combat Bots. The proposed solution adopts a two pronged strategy which we have classified into the standalone algorithm and the network algorithm. The standalone algorithm runs independently on each node of the network. It monitors the active processes on the node and tries to identify Bot processes using parameters such as response time and output to input traffic ratio. If a suspicious process has been identified the network algorithm is triggered. The network algorithm will then analyze conversations to and from the hosts of the network using the transport layer flow records. It then tries to deduce the Bot pattern as well as Bot signatures which can subsequently be used by the standalone algorithm to thwart Bot processes at their very onset.

# 1. Introduction

## 1.1 Background

Commercial as well as governmental organizations are increasingly relying on computer networks to share and process important data. A significant threat to any network is the presence of Botnets. Botnets are considered as compromised computers which can be present anywhere from homes, schools, businesses and even governments around the world. They work under the control of a single hacker, commonly known as a Bot-master. Botnets are often used to conduct attacks ranging from Distributed Denial of Service to corporate intelligence or surveillance and spam delivery. The Botnets have emerged as the number one source of spam over the past years, giving spammers access to virtually unlimited bandwidth.

Spammers do not pay for the messages they send, and hence can e-mail larger documents, like image files (Zhijun Liu et. al., 2005; Fulu Li, Mo-Han Hsieh, 2006; Husain Husna et. al., 2008).

Botnets have come to flood the Internet, largely unnoticed by the public. On a typical day, large number of computers connected to the internet is Bots engaged in distributing e-mail spam, stealing sensitive data typed at banking and shopping websites, bombarding websites as part of extortionist Denial of Service attacks, and spreading fresh infections. IRC bots have been used for infecting hosts by installing malicious code on them (Zhenhua Chi, Zixiang Zhao 2007; Zhijun Liu et. al., 2005). DDoS and other network based attacks render services hosted in a particular network unresponsive and unusable as in case of UDP flood attack. These kinds of attack reduce the throughput of the network as a whole. *Trinoo*, a DDoS attack is capable of compromising thousands of hosts in a given network (David Dittrich, 1999).

## 1.2 Related work

Most of the previous researches done to combat Bots have been centered on the Botnets that follow the C&C (commands & control) model where Bot-masters mainly use the IRC protocol to invoke commands for Bots. In this model a single Bot-master controls all the Bots of a network. Thus, it is not very difficult to thwart the Botnet as only the Bot-master needs to be traced. Unfortunately, most of the recent Botnets have migrated to the P2P (peer to peer) architecture (David Dagon et. al., 2007; Craig A. Schiller, Jim Binkley 2007). This model has a greater capability of regeneration as there is no single point of control and hence it is more dangerous (Ping Wang, Sherri Sparks, Cliff C. Zou 2007). Most of the previous attempts to detect and combat Bots have focused on a limited set of attacks and the analysis for the same were performed based on the presumption that the network is infected by only those attacks. In (Akiyama et al., 2007; Al-Hammadi and Aickelin, 2008; Lei Liu et al., 2008; Stéphane Racine, 2003-04; Wicherski et al., 2005; Georg Wicherski et. al., 2005; Niels Provos, Thorsten Holz 2007), suggested approaches and analysis are based on Botnets that follow C&C model and typically use IRC protocol. Also these approaches provide supervised learning as they try to analyze Bot activities based on certain predefined properties and Bot lifecycles (M.A. Rajab, J. Zarfoss, F. Monrose, A. Terzis 2006). In (Guofei Gu et. al., 2008), the adopted approach tries to analyze traffic patterns generically. However it is a passive approach to detecting Botnets.

Figure 1 below shows the layout of a typical Botnet. We would be using the terminologies mentioned in this figure, *attacker, master* and *agent*, for the remaining sections of the paper.

Figure 1 Representation of a Botnet.

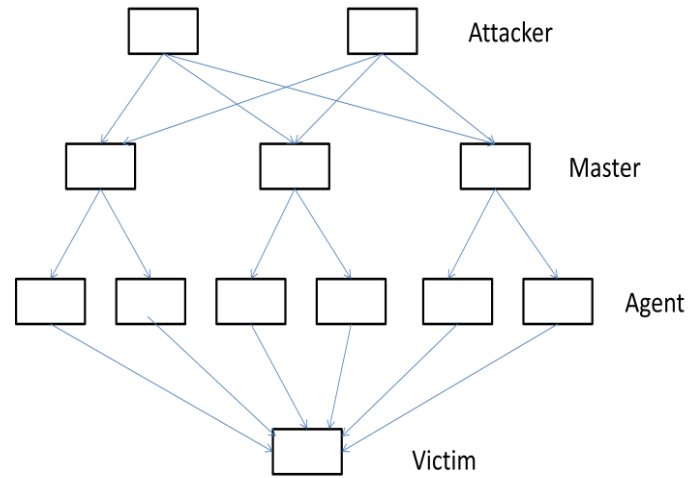

*Attacker*: initiates and controls the attack
*Master*: compromised host that controls the agents and invokes commands
*Agent*: bots that perform P2P attacks on victims based on commands invoked by master

The contribution of our work is as follows:
- A two pronged approach towards detecting Botnets. An algorithm that based on standalone host activity analysis triggers a network analysis to find Botnets in the given network.
- A method for identifying similarities between filtered Bot traffic using Dynamic Time Warping (DTW) algorithm, K-means clustering and graphical analysis.
- We also present an experimental simulation of UDP flood attack and perform analytical calculations on UDP packet flows retrieved from the same, based on the Dynamic Time Warping (DTW) algorithm.

Our solution does not assume the existence of a particular type of Bot and is hence generic. It can deal with both C&C and P2P Bots because of our two pronged approach. The advantage of our solution is that it can evolve to identify new Bot patterns thus making it a learning based approach. The chances of false positives are also reduced because of the two pronged strategy adopted.

The rest of the paper is structured as follows: Section 2 explains the data structure that we use for network traffic flow collection, Sections 3, 4 explain our solution which comprises of Stand Alone and Network Algorithm. These sections also include required experimental analysis. In Section 5 we summarize our work followed by the future work and references in Section 6 and 7 respectively.

## 2. Data Collection and Flow representation

In order to detect malicious and abnormal activities in a network we use a data structure, *flow_data*, to store network activity of each host over a fixed interval of time. For each host, we maintain an array of *flow_data*, *flow_data_array*, such that *flow_data_array[i]* represents the captured *flow_data* in the $i^{th}$ time interval. The *flow_data_array* not only gives us a snapshot of the network activity for a single host but also a combined view of the network as a whole (when *flow_data_array* from all hosts in the network is considered). The data structure definition is as follows:

```
struct address
{
string ip_address;
int port;
}

struct flow_data
{
string  packet_type ;
map<address, long> hostwise_weight_incoming;
map<address, long> hostwise_weight_outgoing;
float avg_response_time;
int icmp_errors;
};
flow_data [] flow_data_array;
```

*hostwise_weight_incoming* represents the mapping of hosts (for a particular port) to incoming traffic volume and *hostwise_weight_outgoing* is used for the host to outgoing traffic volume mapping. Using *flow_data_array* for all hosts in the network, we can construct a weighted (directed) graph to find abnormal, malicious activities in the network. We use the weighted (directed) graph, for network activity analysis, in the network algorithm.

It is important to note that we maintain both *hostwise_weight_incoming* and *hostwise_weight_outgoing* for calculating O/I ratio and for similar analytical calculations in the standalone algorithm. Other parameters included are *packet_type* (TCP, UDP, ICMP, etc.), *icmp_errors* number of ICMP error packets and *avg_response_time*. In the remaining sections we will use *flow_data* to represent the network traffic flow capture and *flow_data_array* to represent the collection of *flow_data* captures over multiple time intervals.

## 3. Stand-Alone Algorithm

As we have mentioned, our solution adopts a two pronged approach. We will refer to the first phase of our solution as the Stand-Alone Algorithm primarily because this algorithm runs independently on each node of the network. The main objective of the algorithm is to monitor active processes on a given node and identify Bot/suspicious processes by evaluating the packets being sent and received by that process along with other relevant parameters such as response time, output to input traffic ratio, APIs invoked etc. In order to simplify the explanation of our algorithm we have established various states that a monitored process might be in as the algorithm progresses. The algorithm begins as soon as the computer boots with all scheduled processes being in the initial state. Once a process begins execution, it enters the execution state and we begin to monitor its activities. If we find the activities to be suspicious, the process is updated for its status as "SUSPECT" and then moved to a specially designed task manager, which assigns the process a lower priority and keeps a tab on the system resources that the process is utilizing. However, normal processes will continue to remain in the execution state and will eventually migrate to the terminate state when the processes end. For each suspicious process, a suspicion value is calculated. If this value exceeds our thresh-hold value, the network algorithm is triggered. We will now explain the methodology for calculating the suspicion value and the rationale behind it.

We calculate the suspicion value based on certain parameters which are as follows:

Response Time: This is the elapsed time between the submission of a request and the beginning of a response from a particular host. The analysis of this parameter is based on the type of attack, as follows:
*IRC Bot*: IRC Bots are developed using an event based framework wherein Bots invoke commands like MSG, ACTION, ECHO and JOIN in response to events/actions like TEXT, JOIN, CHAT, and PART. IRC Bots are programmed to respond to events/actions either instantaneously or after a fixed amount of time. Events/actions for Bots include a user entering a text (TEXT) or a user joining an IRC channel (JOIN), response to which can be; displaying a message on the IRC channel or performing some other actions depending on the event.

A simple IRC Bot script is as follows:
```
on 1 : TEXT : Hello : #mIRC : {
        /msg $chan Hello.....!
}
```
In such a scenario, the response time between an incoming IRC packet (a comment/event packet to notify an activity in the channel) and an outgoing IRC response packet is very small. A simulated IRC Bot execution showed that the time gap between an incoming IRC event request and the corresponding response is ~221 ms.
*DDoS attack*: Network based attacks like DDoS attack are organized in terms of the architecture of the various compromised hosts involved in the attack as shown in Figure 1. The response time between an attacker issuing a command to the master and the master responding to the same is very less (in the order of milliseconds). Similar is the case between the master and the agents. Other network based attacks like Ping of Death, Smurf attack and Spam attack follow similar traits as far as response time is concerned.

We keep track of the average response time for individual hosts to identify potentially compromised hosts (Bots) with low values for response time (as compared to

221ms). This can be achieved by evaluating *avg_response_time* field of *flow_data* for individual hosts.

IP Address: We check the source IP address of incoming packets and the destination IP address of outgoing packets to determine whether any of these IP address have previously been blacklisted by the network algorithm due to suspicious activities at the corresponding nodes. We also filter out IP addresses of legitimate hosts like servers hosting standard service in the network (Guofei Gu et. al., 2008).

Network Traffic Pattern: Bots engaging in different type of attacks such as spam attack, DDoS attack and key logging attack produce network traffic following a particular pattern (James R. Binkley, Suresh Singh 2006; Guofei Gu, Junjie Zhang, and Wenke Lee 2008; Yousof Al-Hammadi and Uwe Aickelin 2008; Jae-Seo Lee et.al., 2008; Anestis Karasaridis, Brian Rexroad, David Hoeflin 2007; Mitsuaki Akiyama et. al., 2007).

We use the following characteristics for determining a malicious network traffic pattern:
- Consecutive packets sent by the Bot process having identical values in the protocol header fields. DDoS attack for example involves a flood of packets having the same destination IP address, protocol (TCP, UDP, ICMP or IRC based on the type of attack the Bot is attempting) and protocol header field values.
- Intermittent peaks in the network IO graph for hosts in the network.
- High volume of outgoing TCP, UDP packets in response to minimal but controlled incoming network traffic.

A simulated execution of an IRC Bot showed controlled outgoing IRC traffic only in response to certain specific incoming IRC packets with a very small and almost constant response time.

Figure 2 IRC traffic flow capture for IRC Bot.

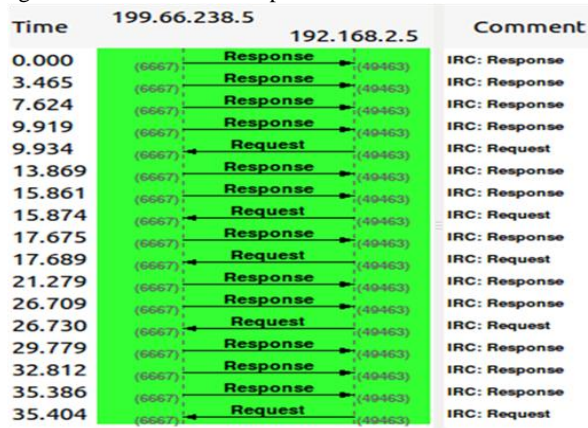

A snapshot of the IRC traffic flow captured is shown in Figure 2, above.

The time specified in the above figure has the following format:
<packet-sequence>.<time in milliseconds>

Entries with same packet-sequence value represent the following sequence:
- Notification from the IRC server to the IRC Bot for an action in the IRC channel.
- A request initiated from the IRC Bot in response to the notification received from the IRC server.

It is important to note that these patterns are instrumental to not only detect Bots but also to detect the Bot-master (attacker). We keep a track of hosts following the below mentioned patterns:
- A low response time
- High volume of TCP SYN packets
- High volume of TCP, UDP packets
- Regular TCP, UDP packets having Bot commands, as mentioned in (David Dittrich, 1999), in the payload.

Ports Used for Communication: Most Bots use particular ports as endpoints of communication. These ports can give an indication of the kind of attack a Bot might be engaging in because typically Bots use different ports for different type of attacks. For example IRC Bots use port 6667.

For network oriented attacks like DDoS, Spam, Tear Drop and Smurf attack there are multiple ports that are used depending on whether the conversation is between the *attacker and master* or *master and agent* or *agent and victim*. And the ports used depend on the results of the network scanning phase.

We keep track of the traffic flow on standard Bot ports and also keep track of open ports on network hosts that can be potentially used by Bot-masters to invoke commands to Bots.

Continuous Attempts for Connection Setups: As Bots try to propagate across the network, Bot processes need to continuously establish connections with other nodes in the network. Most of the IRC chat servers keep a flooding limit for participants on the channel. This leads to frequent disconnections from IRC channels for IRC Bots and subsequent connection attempts.

A DDoS attack also involves a lot of connection attempts and failures during the network scanning phase. Thus, by keeping a tab on the frequency at which connection attempts are being made by a process, we can probabilistically determine if the concerned process is a Bot process using (*Ourmon - network monitoring and anomaly detection system*). We also keep track of *UDP work weight*, (Binkley and Parekh, 2009) for potentially compromised hosts. *UDP work weight* roughly measures the amount of network noise caused by a host (Binkley and Parekh, 2009). A high value for UDP work weight is indicative of a Bot host. *UDP work weight* calculation for a simulated UDP flood attack for one of the compromised host is as follows:

SENT (UDP packets sent from the host) = 2649 packets
RECV (UDP packets received on the host) = 561 packets
ICMPErrors (Incoming ICMP error packets) = 1927 packets

UDP work weight =
(SENT x ICMPErrors) + RECV = 5105184    (Binkley and Parekh, 2009)

High *UDP work weight* in this case indicates a host trying to scan a large network as fast as possible or involved in a DDoS attack.

Number of Active Connection: A Bot process typically has a high number of active TCP connections through which it sends and receives packets. Bot-masters/attackers use these active connections to invoke commands, as mentioned in (David Dittrich, 1999), on compromised hosts (Agents) to perform certain actions. Hence it is an important parameter in determining the suspicion value. A high value for *hostwise_weight_incoming* on Bots, victim hosts and a high value for *hostwise_weight_outgoing* on attacker hosts indicate a potential Bot activity.

Output to Input Traffic Ratio (O/I Ratio): This is the ratio of the total outgoing traffic from a process to the total incoming traffic for that particular process. This parameter is useful in identifying the kind of attack a Bot process might be engaging in. For example, the O/I Ratio for Bots involved in spam or DDoS attack is very high while for key-logging Bots the O/I Ratio is very low.
Snapshot of network activity during the infection phase of a simulated DDoS attack is as below:

Figure 3 Outgoing TCP traffic from one of the Bots/attacker involved in DDoS attack.

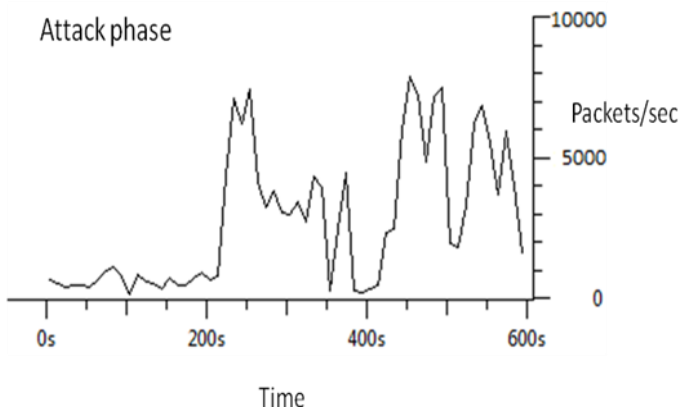

Thus a high O/I ratio for a host is indicative of Bots activity. This again can be calculated using the *hostwise_weight_incoming* and *hostwise_weight_outgoing* fields of *flow_data*.

It must be noted that these parameters are applied to the incoming as well as outgoing traffic of a process. Moreover, not only do these parameter help us determine whether a process is suspicious or not but they also help us classify the suspicious process into one of the three categories, based on the possible kind of attack that the process is carrying out. These categories are Key-logging, DDoS and Spam. The suspicion value is a weighted average of the computed numeric values for the above parameters. In the next phase, the Stand-Alone Algorithm carries out further analyses of the suspicious processes based on the category into which they have been classified. The three possible analyses are as follows:

- For a process which has been classified into the key-logging category, we start monitoring the API functions that the process is invoking. We are mainly concerned with three kinds of functions which are; communication functions, file access functions and keyboard-state functions. The objective of a key-logging Bot is to keep a tab on all the keys being pressed on the target machine and convey this information to the Bot-master. We use Spearman's rank correlation to conclusively determine whether the process is a key-logging Bot or not (Yousof Al-Hammadi and Uwe Aickelin, 2008). Spearman's rank correlation, used in statistical analysis, assesses how well the relationship between two variables can be described using a monotonic function (*Spearman's rank correlation coefficient*). We determine the spearman rank correlation coefficient amongst the frequency of invocations of the three types of API functions mentioned above.
- For a process which has been classified into the DDoS category, we start evaluating the process activities in order to identify the kind of DDoS attack. For instance, in case of the Ping of Death Attack the Bot will generate ping packets which are exceptionally large in size. Thus we can detect this attack by checking the size of the outgoing ping packets. The other kinds of DDoS attacks that the Stand-Alone Algorithm can detect are Flood Attack, SYN attacks, Tear Drop Attack and Smurf Attack. The details of detection have been omitted here for the sake of brevity.
- For a process which has been classified into the SPAM category we evaluate its traffic characteristics. We perform a contextual analysis which scans the message body in order to identify certain markers which are known to be associated with spam. Similarly, we also check for message similarity across consecutive messages. Based on this analysis, we try to conclusively determine whether the suspicious process is really engaging in a spam attack or not.

In the last stage of the Stand-Alone Algorithm, we perform the Process Log Analysis. This analysis is based only on the kind of APIs a process is invoking and their corresponding timestamp. The objective is to evaluate the type and frequency of the various API functions invoked by a process during its lifetime in order to identify a definitive pattern which will help us to determine not only whether the process is a Bot or not but also the kind of attack it may be engaging in.

##  Network Algorithm

The main objective of the Network Algorithm is to function as a central algorithm that examines the transport layer traffic, *flow_data*, captured from the entire enterprise network and try to find useful patterns of Bot behavior which can then be used by the individual nodes to detect Bot attacks with greater accuracy and efficiency. This algorithm is passive in nature and is activated by alerts from the Stand-Alone Algorithm. The various stages involved are as follows:

Reception of Trigger Events: The algorithm receives triggers from the Stand-Alone Algorithm that conveys information such as originator machine IP address, Bot process detected, inbound and outbound ports and IP addresses of the infected hosts and the type of attack detected by Stand-Alone Algorithm.

Identification of Machines with Suspected Behavior and Selection of Flow Records: Based on these triggers, the machines sending the trigger and remote host at the Inbound IP addresses are shortlisted as the machines with suspected behavior. This is followed by extraction of the subset from *flow_data* containing either of these IP addresses in the Source IP or Destination IP fields. This can be done by retrieving the information from the *flow_data* array for each host and performing search on the *flow_data* for the IP addresses.

Identification of Suspected Conversations: To eliminate legitimate traffic, a two step process is used which consists of:

- Packet Type Filtering: We select only those flow records in which the Packet Type is IRC, HTTP, TCP, UDP, and ICMP, as these are the protocols which are used by Bots and are indicative of a malicious activity. The *type* field in the *flow_data* indicates what type of traffic (TCP, UCP etc) we want to capture.

- Port Filtering: Port based filtering is possible only in case of Bots that operate on standard ports, for example 6667, 6668, 7000/tcp, (Karasaridis, Rexroad and Hoeflin, 2007;Wang, Sparks and Zou, 2007; Wicherskiat al., 2005). It is important to note that for most of the network based attacks like DDoS, the ports used for command invocation, controlling and the actual attack are not standard. They are dependent on the results of the network scanning phase. Moreover attackers use different ports for communication between

    o master and agent

    o attacker and master

In fact, even the protocols used for the two conversations are different. The primary reason behind this is to avoid any correlations in port usage, (David Dittrich, 1999). For non-standard cases, port filtering can however be done based on further analysis of the malicious *flow_data* captured.

To detect Bots using non-standard ports:

o We use the property that Bot controllers usually have one to many port connections with their targets, (Karasaridis, Rexroad and Hoeflin, 2007).

o We find flow records between the suspected Bots and remote servers which have traffic characteristics within the bounds of a flow model for IRC traffic, (Karasaridis, Rexroad and Hoeflin, 2007).

Filtration of Suspected Conversations: From the identified suspect conversations we will obtain the most likely Bot conversations (command as well as activity conversations) using the following techniques in parallel.

- Response time clustering: It has been observed that the response time between receptions of commands from Bot-master and initiation of malicious activity is substantially small compared to typical human IRC responses. Using this principle, we find response times for all suspected nodes in the network. We use K-means clustering algorithm, to partition nodes having low response time. These partitioned nodes are considered for next filter phase. K-means clustering algorithm partitions *n* observations into *k* clusters in which each observation belongs to the cluster with the nearest mean.*(K-means clustering)*

- Synchronization filtering: It has been observed that on reception of commands from same Bot-master all the zombie hosts initiate malicious activity at roughly the same time. Using this principle we consider all activity conversations that started at the similar time for the next phase. This can be achieved by analyzing *flow_data* snapshots, captured for all hosts in the $n^{th}$ time interval, for simultaneous abnormal activities.

Analysis of filtered network traffic using Dynamic time warping: Bots involved in DDoS attacks like UDP, TCP or ICMP flood, produce network traffic having similar patterns. The challenge lies in trying to find a similarity score for such traffics. We use Dynamic Time Warping (DTW) algorithm, often used in speech recognition techniques, to find similarities in network traffic patterns generated by different hosts in a network. Dynamic time warping (DTW), is an algorithm that measures the similarity between two sequences which may vary in time or speed *(Dynamic time warping; Elena Tsiporkova, Dynamic time warping)*. The rationale behind using this algorithm is as follows:

- The sequences that can be compared using this algorithm may vary in time and speed.

- The algorithm provides a non-linear (elastic) alignment, which produces a more intuitive similarity measure, allowing similar network traffic patterns to match even if they are out of phase across the time axis (*Elena Tsiporkova, Dynamic time warping* ).
- The kind of sequences that DTW algorithm can analyze are similar to network traffic sequences we intent to analyze.

The formula to calculate the similarity score for two sequences A, B having m, n data points respectively is as follows :

$$D(A, B) = \frac{\sum_{s=1}^{k} d(P_s) * W_s}{\sum_{s=1}^{k} W_s}$$

*(Elena Tsiporkova, Dynamic time warping)*

$D$: distance between sequences A and B
$d(P_s)$: distance between $i_s$ and $j_s$.
$P$ is the function representing points across the optimized (least distance) path between the two sequences.
$w_s > 0$: weighting coefficient
We use the weighting coefficient such that

$$C = \sum_{s=1}^{k} W_s$$

Here $C = n + m$ as we use the symmetric form for the weighting coefficient

$$D(A,B) = 1/C \sum_{s=1}^{k} d(P_s) * W_s$$

$D(A, B) = g(n, m) / C$

We calculate the optimized value for $D$ (A, B) using dynamic programming. Dynamic programming is a method for solving complex problems by breaking them down into simpler sub problems *(Dynamic programming)*. We apply Dynamic programming in our calculation as follows:

Initial condition: $g(1, 1) = 2d(1, 1)$.

$$g(i,j) = min \begin{Bmatrix} g(i,j-1) + d(i,j) \\ g(i-1,j-1) + 2d(i,j) \\ g(i-1,j) + d(i,f) \end{Bmatrix}$$

$g(i,j)$ : min. value of function P at point (i,j)
$d(i,j)$ : Euclidian distance between point i of sequence1 and point j of sequence2.

Thus if $D$ (A, B) < $D$ (B, C) then the sequences A and B are more similar as compared to sequences B and C. We have used 'packets/sec' as the metric, in this paper, for our experiments and analytical computations. It is important to note that the 'network traffic pattern analysis' step in the standalone algorithm compares network traffic generated with well-known and archived malicious network traffic patterns. In this step however, we analyze network traffic generated by hosts at run time and compare them with each other and not with a standard pattern.

Analysis of UDP traffic from two Bots performing UDP flood attack and an uninfected host having normal UDP traffic is as below. We have used the R-project's DTW package (Toni Giorgino, 2009) to compute the similarity scores i.e. $D$ (A, B). We have used Wireshark *(Wireshark)* to capture UDP packets and to generate IO graphs for the same. We have considered 'packets/sec' captured from hosts at time steps of 5 seconds.

Seq1: Bot involved in UDP flood attack.
Seq2: Bot involved in UDP flood attack.
Seq3: Normal uninfected host

Analysis of Seq1 and Seq2:
Seq1: [123, 2387, 2265, 2465, 2409, 2057, 257, 1334, 511, 2343, 2426, 2489, 2412, 1324, 784, 1213, 2334, 213, 734, 755]

Seq2: [130, 153, 514, 2414, 2389, 2337, 2214, 2034, 1807, 1753, 1764, 2419, 2309, 2409, 2411, 2216, 2367, 2354, 1302, 456, 2456, 1215, 1347, 1385, 124]

Figure 4 Seq1, Bot involved in UDP flood attack

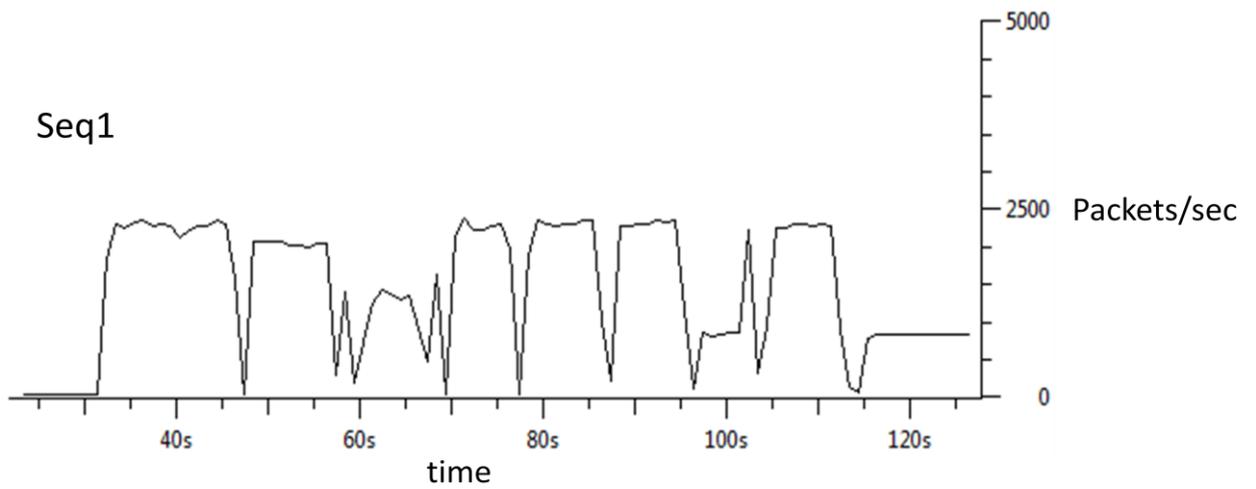

Figure 5 Seq2, Bot involved in UDP flood attack

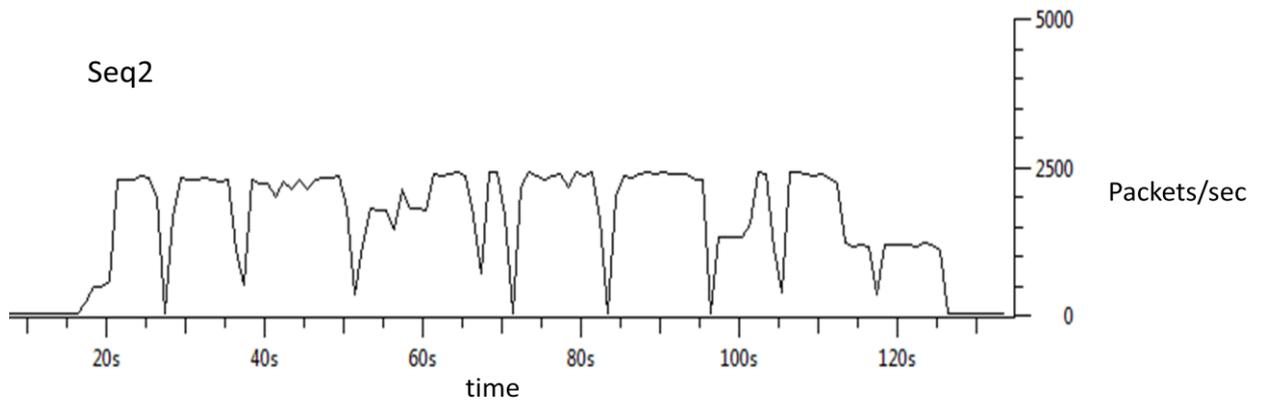

$D$ (Seq1, Seq2) = 8729

Analysis of Seq1 and Seq3:

Seq1: [123, 2387, 2265, 2465, 2409, 2057, 257, 1334, 511, 2343, 2426, 2489, 2412, 1324, 784, 1213, 2334, 213, 734, 755]

Seq3: [11, 33, 25, 27, 103, 123, 124, 29, 63, 9, 52, 51, 53, 48, 23, 35, 33]

Figure 6 Seq3, Normal (legitimate) UDP traffic

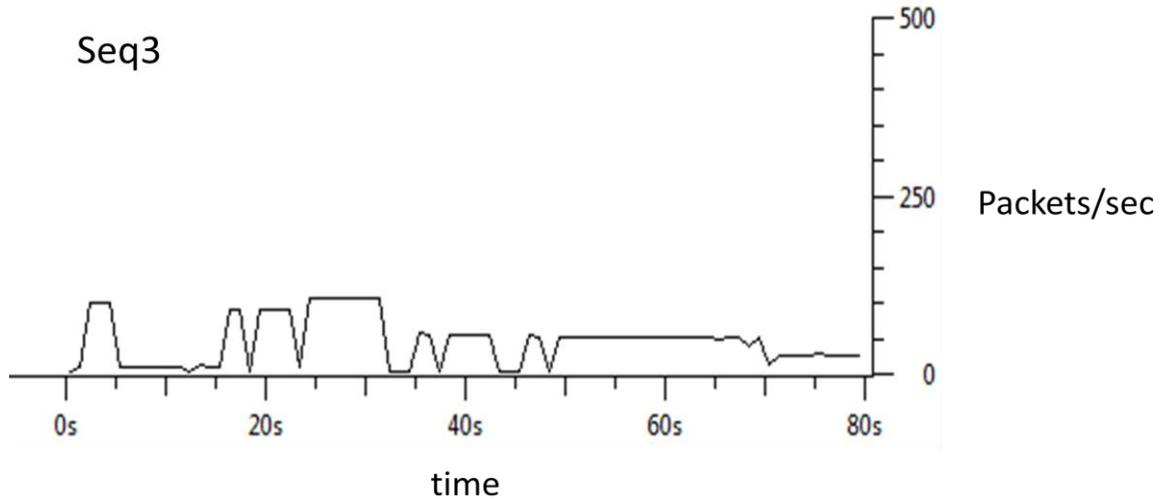

$D$ (Seq1, Seq3) = 30689

In the above analysis $D$ (Seq1, Seq2) < $D$ (Seq1, Seq3), thus implying that Seq1 and Seq2 are more similar as compared to Seq3. This observation supports our proposition that 'Bots involved in a particular type of attack produce similar network traffic pattern', since Seq1 and Seq2 are generated by Bots involved in UDP flood attack and Seq3 is generated by a normal uninfected host.

Analysis of filtered conversations using graphical analysis and clustering: The conversations obtained from the union of results of the two filtration techniques will be analyzed to detect groups of compromised machines with similar communication patterns and similar malicious activity patterns. In this phase we also attempt to classify the type of malicious activity and finally generate a set of Bot signatures as well as soft black list. This phase consists of following three activities:

- Graphical Analysis: We analyze the filtered conversations by constructing a weighted (directed) graph to represent the network traffic for all the hosts in the network for the n$^{th}$ time interval. Nodes in the graph represent hosts in the network and edges represent network traffic between them. The weighted (directed) graph can be constructed by considering *hostwise_weight_incoming* and *hostwise_weight_outgoing* of the n$^{th}$ *flow_data* in *flow_data_array* for each of the hosts in the network. Figure 7 shows a weighted (directed) graph constructed in a similar way.

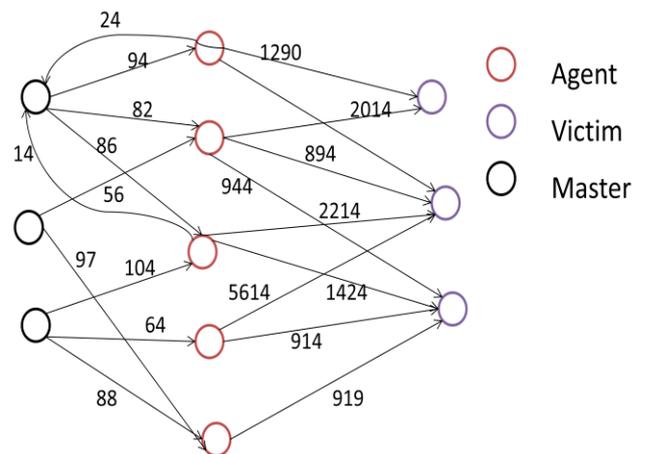

Figure 7 A weighted (directed) graph constructed using *flow_data* captures

In the above graph, weights assigned to edges are the TCP (or UDP) packet volume (outgoing or incoming based on the direction) for the hosts. The graph shows high weight values for outgoing edges emerging from some of the nodes (agents) and low values for incoming edges ending at them. This is indicative of a Bot carrying out a DDoS attack based on the commands invoked from its Bot-master. We identify the nodes:

- That have a high overall outgoing weight and low incoming weights.

- That has incoming edges that correspond to TCP and UDP traffic carrying Bot commands (David Dittrich, 1999).

It should be noted that the graph constructed above is not based on any simulations or real time network

statistics and is hence not conclusive. It is used to demonstrate:
1. How to construct a network flow weighted graph based on the *flow_data* captured.
2. How network traffic patterns like high O/I ratio concentrated in some regions of the network can be deduced from this graph.
3. How the graph can be used to check if there is a high volume of outgoing malicious traffic from compromised hosts at about the same time.

- Clustering:

*Command dimension clustering*: This step mainly concentrates on the traffic flow data i.e. *who is talking to whom?* We filter out:

- All flows that are directed from internal to external hosts.
- Communications between internal hosts as well as communications, between hosts, that are not completely established (that contain only one way traffic).
- Flows to known legitimate servers.

We then use the following parameters to cluster the conversations (Guofei Gu et. al., 2008):

- Number of flows per hour (fph)
- Number of packets per flow(ppf)
- Average number of bytes per packet(bpp)
- Average number of bytes per second(bps)
- Port number

*Activity dimension clustering*: This step mainly concentrates on the activities that the hosts are involved in i.e. *what are the hosts doing?* We use a two layer clustering scheme. In the first layer, we cluster the conversations according to type of malicious activity. In the second layer we cluster the conversations according to activity features. The types of activities and the specific cluster features for each of them are as given below (Guofei Gu et. al., 2008):

- Port scan activity: Features include rigorous attempts by Bots to find potential vulnerable ports on hosts within the same subnet. It also leads to high volume of failed connection attempts.
- Spam activity: Two clients would be clustered together if they are using same SMTP connection destination.
- Binary downloading activity: Features could include similar binary, similar URL, etc.
- DDoS: Features include high bytes per packet, large number of flows per connection and high number of packets per sec

*Cross dimension correlation*: We select one cluster from the command dimension and one cluster from the activity dimension and compute the cross correlation. This process is repeated for each pair of clusters in the command and activity dimensions. From these, we select those pairs that have high correlation. (Guofei Gu et. al., 2008).

Issue of alerts: Based on the analysis of the standalone and network algorithm an alert is issued to take necessary measures. Also an alarm is issued to all hosts to download the latest signatures.

## Summary

In this paper we have proposed a novel two pronged approach to detect Botnets in a given network. We proposed a data structure that efficiently stores network flow data. This data structure is used for analysis in the standalone and network algorithm. The standalone algorithm is heuristic in nature whereas the network algorithm relies on network traffic analysis. The two pronged approach used in this paper helps to analyze the network captures effectively because the intensive network analysis, that require more effort is triggered only when standalone algorithm generates an alarm. Because of the two pronged approach, adopted the chances of false positives are reduced as the network algorithm is triggered only if the standalone algorithm raises an alarm. The analysis of network traffic flow, based on Dynamic Time Warping (DTW) algorithm performed on experimental values showed positive results for similarity in Bot traffic patterns.

## Future Work

In future we intend to focus our efforts towards the following:
- Implement a framework that effectively applies the suggested two pronged strategy to large scale real-time enterprise networks. We also intend to make this framework generic in terms of the parameters used to evaluate network activities as well as individual host activities. The main motive behind this is to ensure that this framework can cope up with new Bots and Botnet models that may evolve in the future.
- Make our two pronged approach real time by using the Reval tool, suggested by (Vasudevan et al., 2006).

- Reval is an operational support tool, used to mitigate the impact of network based attacks like DDoS. It can scale to large networks and detect abnormal network activities in real time.
- The kind of network traffic flow data that is used by the network algorithm for network flow analysis is large, typically in gigabytes. Running the network algorithm on such huge volumes of data effectively is a challenge. We intend to apply various optimization techniques to efficiently analyze the network flow data.
- Estimate the complexity of uploading the *flow_data* array to a specific node and computation complexity of extracting the occurrence of an attack in real time, using the Reval tool, (Vasudevan et al., 2006) to establish the feasibility of real-time detection.

## References

Georg Wicherski, Markus Kötter, Paul Bächer, Thorsten Holz (2005). 'Know your enemy: Tracking Botnets'. *The Honeynet Project & Research Alliance*. Available at http://www.honeynet.org/papers/bots/ (Accessed 6[th] November, 2011), pp.1-17.

Anestis Karasaridis, Brian Rexroad, David Hoeflin (2007) 'Wide-scale Botnet Detection and Characterization'. *HotBots'07 Proceedings of the first conference on First Workshop on Hot Topics in Understanding Botnets*, USENIX Association Berkeley, CA, USA, pp.7-7.

Niels Provos, Thorsten Holz (2007) 'Virtual Honeypots: From Botnet Tracking to Intrusion Detection' 1[st] ed. Addison-Wesley Professional, ISBN: 9780321336323.

M.A. Rajab, J. Zarfoss, F. Monrose, A. Terzis (2006) 'A multifaceted approach to understanding the Botnet phenomenon'. IMC '06 *Proceedings of the 6th ACM SIGCOMM conference on Internet measurement*, New York, NY, USA. pp. 41-52. ISBN: 1-59593-561-4, DOI: 10.1145/1177080.1177086.

Guofei Gu, Roberto Perdisci, Junjie Zhang, and Wenke Lee (2008) 'BotMiner: Clustering Analysis of Network Traffic for Protocol- and Structure-Independent Botnet Detection'.*SS'08 Proceedings of the 17th conference on Security symposium*, July 28-August 01, 2008, San Jose, CA, USA, pp. 139-154.

Al-Hammadi, Yousof and Aickelin, Uwe (2006) 'Detecting Botnets through Log Correlation*'*. In: *Proceedings of MonAM 2006 - IEEE/IST Workshop on Monitoring, Attack Detection and Mitigation*. Tuebingen, Germany, (September 28–29, 2006), pp. 97-100.

James R. Binkley, Suresh Singh (2006) 'An Algorithm for Anomaly-based Botnet Detection'. *SRUTI'06 Proceedings of the 2nd conference on Steps to Reducing Unwanted Traffic on the Internet* - Volume 2, USENIX Association Berkeley, CA, USA, pp. 43-48.

Mitsuaki Akiyama, Takanori Kawamoto, Masayoshi Shimamura, Teruaki Yokoyama, Youki Kadobayashi, Suguru Yamaguchi (2007) 'A proposal of metrics for botnet detection based on its cooperative behavior'. *SAINT-W '07 Proceedings of the 2007 International Symposium on Applications and the Internet Workshops,* IEEE Computer Society Washington, DC, USA, pp. 82-82, ISBN: 0-7695-2757-4, DOI: 10.1109/SAINT-W.2007.14.

Yousof Al-Hammadi and Uwe Aickelin (2008) 'Detecting Botnets Based on Key logging Activities'. *Proceedings of the Third International Conference on Availability Reliability and Security,* (ARES 2008), pp. 896-902. Barcelona, Spain. Publisher: IEEE Computer Society.

David Dagon, Guofei Gu, Christopher P. Lee, Wenke Lee, Georgia Inst. of Technol., Atlanta (2007) 'A Taxonomy of Botnet Structures'. In *Proceedings of the 23 Annual Computer Security Applications Conference* ACSAC'07 (2007), pp. 325-339 ISBN**:** 978-0-7695-3060-4.

Ping Wang, Sherri Sparks, Cliff C. Zou (2007) 'An Advanced Hybrid Peer-to-Peer Botnet'. *IEEE Transactions on Dependable and Secure Computing*, 7(2), pp. 113-127, April-June, 2010.

Fulu Li, Mo-Han Hsieh (2006) 'An Empirical Study of Clustering Behavior of Spammers and Group-based Anti-Spam Strategies'. In Proceedings of *CEAS 2006 – Third Conference on Email and Anti-Spam*, July 27-28, 2006, Mountain View, California USA

Stéphane Racine (October 2003 - April 2004) 'Analysis of Internet Relay Chat Usage by DDoS Zombies', Master's Thesis, MA-2004-01, Swiss Federal Institute of Technology, Zurich. Available at: ftp://ftp.tik.ee.ethz.ch/pub/students/2003-2004-Wi/MA-2004-01.pdf

Husain Husna, Santi Phithakkitnukoon, Srikanth Palla, and Ram Dantu (2008) 'Behavior Analysis of Spam Botnets'. Communication Systems Software and Middleware and Workshops, 2008. COMSWARE 2008. 3rd International Conference on 6-10 Jan. 2008, pp. 246-253, ISBN**:** 978-1-4244-1796-4, Issue Date**:** 6-10 Jan. 2008.

Craig A. Schiller, Jim Binkley (2007). 'Botnets the Killer Web App'. Syngress Publishing, ISBN-10: 1597491357 | ISBN-13: 978-1597491358. Publication Date: February 15, 2007.

Lei Liu, Songqing Chen, Guanhua Yan, and Zhao Zhang (2008) 'BotTracer: Execution-based Bot-like Malware Detection'. ISC '08 Proceedings of the 11th international conference on Information Security Springer-Verlag Berlin, Heidelberg (2008), pp. 97-113. ISBN: 978-3-540-85884-3, DOI: 10.1007/978-3-540-85886-7_7.

Guofei Gu, Junjie Zhang, and Wenke Lee (2008) 'BotSniffer: Detecting Botnet Command and Control Channels in Network Traffic'. In *Proceedings of the 15th Annual Network and Distributed System Security Symposium (NDSS'08)*, San Diego, CA, February 2008.

Zhenhua Chi, Zixiang Zhao (2007) 'Detecting and Blocking Malicious Traffic Caused by IRC Protocol Based Botnets'. *NPC '07 Proceedings of the 2007 IFIP International Conference on Network and Parallel Computing Workshops* IEEE Computer Society Washington, DC, USA (2007), pp. 485-489. ISBN: 0-7695-2943-7.

Zhijun Liu, Weili Lin, Na Li, and David Lee(2005) 'Detecting and Filtering Instant Messaging Spam – A Global and Personalized Approach', *NPSEC'05 Proceedings of the First international conference on Secure network protocols* IEEE Computer Society Washington, DC, USA (2005), pp. 19-24. ISBN: 0-7803-9427-5.

*Dynamic time warping* [online] Available at http://en.wikipedia.org/wiki/Dynamic_time_warping (Accessed 29th October 2011).

Jae-Seo Lee, HyunCheol Jeong, Jun-Hyung Park, Minsoo Kim, Bong-Nam Noh (2008) 'The Activity Analysis of Malicious HTTP-based Botnets using Degree of Periodic Repeatability'. *SECTECH '08 Proceedings of the 2008 International Conference on Security Technology* IEEE Computer Society Washington, DC, USA (2008), pp. 83-86. ISBN: 978-0-7695-3486-2, DOI: 10.1109/SecTech.2008.52.

James R. Binkley, Divya Parekh (2009). *Traffic Analysis of UDP-based Flows in Ourmon*, Portland State University , Computer Science Technical Report – 0807, Computer Science Dept, Portland State University ,Portland, OR,USA, Available at 'http://web.cecs.pdx.edu/~jrb/jrb.papers/flocon2009/flocon2009.pdf' (Accessed 7[th] November 2011).

*K-means clustering* [online] Available at http://en.wikipedia.org/wiki/K-means_clustering (Accessed 12th September 2011).

Toni Giorgino (2009). Computing and Visualizing 'Dynamic Time Warping Alignments in R: The dtw Package'. Journal of Statistical Software, 31(7), pp.1-24. http://www.jstatsoft.org/v31/i07/ (Accessed 1[st] November 2011)

*Wireshark* [online] Available at http://www.wireshark.org/ (Accessed 25[th] October 2011)

Elena Tsiporkova, *Dynamic Time Warping (DTW) Algorithm* [online] Available at www.psb.ugent.be/cbd/papers/gentxwarper/DTWAlgorithm.ppt (Accessed 3[rd] November 2011)

*Dynamic programming* [online] Available at http://en.wikipedia.org/wiki/Dynamic_programming (Accessed 5[th] November 2011).

*Ourmon network monitoring and anomaly detection system* [online] Available at http://ourmon.sourceforge.net/ (Accessed 2[nd] January 2009)

David Dittrich (dittrich@cac.washington.edu), University of Washington (1999). *The DoS Project's 'trinoo' distributed denial of service attack tool* [online] Available at http://staff.washington.edu/dittrich/misc/trinoo.analysis.txt (Accessed 2[nd] November 2011).

*Spearman's rank correlation coefficient* [online] Available at http://en.wikipedia.org/wiki/Spearman%27s_rank_correlation_coefficient (Accessed 20[th] December, 2008.

Rangarajan Vasudevan, Z. Morley Mao, Oliver Spatscheck , Jacobus van der Merwe(2006) 'Reval: a tool for real-time evaluation of DDoS mitigation strategies', *Proceedings of the annual conference on* USENIX '06 Annual Technical Conference, May 30-June 03, 2006, Boston, MA, pp.15-15.